\documentclass[aps,prb,twocolumn,secnumarabic,nofootinbib,superscriptaddress,showpacs]{revtex4-1}

\usepackage{graphicx}
\usepackage{amssymb,amsmath}
\usepackage{color}
\usepackage[usenames,dvipsnames]{xcolor}
\usepackage[colorlinks=true,linkcolor=Red,citecolor=Green,linktoc=page]{hyperref}
\begin{document}

\title{Hotspots in two-phase conducting media}

\author{I.\,A.\,Sadovskyy}
\affiliation{Materials Science Division, Argonne National Laboratory, 9700 S. Cass Avenue, Argonne, Illinois 60637, USA}
\author{A.\,Glatz}
\affiliation{Materials Science Division, Argonne National Laboratory, 9700 S. Cass Avenue, Argonne, Illinois 60637, USA}
\affiliation{Department of Physics, Northern Illinois University, DeKalb, Illinois 60115, USA}
\author{V.\,M.\,Vinokur}
\affiliation{Materials Science Division, Argonne National Laboratory, 9700 S. Cass Avenue, Argonne, Illinois 60637, USA}
\author{P.\,N.\,Kropotin}
\affiliation{A.\,V.\,Rzhanov Institute of Semiconductor Physics SB RAS, 13 Lavrentjev Avenue, Novosibirsk, 630090, Russia}
\author{T.\,I.\,Baturina}
\affiliation{Materials Science Division, Argonne National Laboratory, 9700 S. Cass Avenue, Argonne, Illinois 60637, USA}
\affiliation{A.\,V.\,Rzhanov Institute of Semiconductor Physics SB RAS, 13 Lavrentjev Avenue, Novosibirsk, 630090, Russia}
\affiliation{Department of Physics, Novosibirsk State University, 2 Pirogova Str., Novosibirsk, 630090, Russia}

\date{\today}

\begin{abstract}
We study electric properties of random resistor networks consisting of resistors of two kinds numerically, focusing on the power loss across each bond. Tuning the ratio of the resistances $r$ and their respective fraction $\alpha$ we find that at large $r$ the conductance of the network is dominated by a few optimal, percolation-like, conducting paths. We demonstrate that the distribution of the local power losses $P$ is exponential, $\propto\exp(-P/\langle P\rangle)$, and reveal the spatial distribution of hotspots concentrating the main part of the dissipated power.
\end{abstract}


\maketitle

\section{Introduction
\label{sec:introduction}}

A wide variety of electronic systems that are expected to be spatially homogeneous by their structural properties evolve textures comprising distinct phases differing by their electric properties. Emergence of spatial inhomogeneity is associated with the simultaneous existence and competition of different interactions\cite{Dagotto:2005} such as quantum electron correlations, Coulomb forces, spin, electron-phonon, or even elastic.\cite{Glatz:2011} Archetypal systems of this kind are high temperature superconductors and systems undergoing metal- and/or superconductor-insulator transition (MIT and SIT).\cite{Imada:1998,Levi:2000,Cren:2001,Pan:2001} The latter are characterized by separated competing metal or superconductor and insulating phases.\cite{Kowal:1994} The peculiarity of these systems defining their utmost importance from both, fundamental and applied viewpoints, is their giant responses to small perturbations. One exemplary phenomenon, which is key towards the understanding of metal-insulator- and superconductor-insulator transitions are current jumps in the current-voltage characteristics at some threshold voltage.\cite{Ladieu:1996} The mechanism of the jumps that mark an abrupt transition from a high-resistance insulating to a low-resistance metallic state is thought to be an overheating of the electronic system.\cite{Basko:2007,Altshuler:2009} However, a standard overheating description based solely on the averaged macroscopic material characteristics\cite{Gurevich:1996} cannot explain the behavior of the observed current-voltage characteristics.\cite{Ladieu:1996} It has been long known\cite{Soderberg:1987} that in an inhomogeneous material the current and power loss densities may have strong spatial variations and that areas with enhanced Joule heat dissipation, as compared to their neighboring vicinity, appear. These areas are called {\it hotspots} and may play the role of nuclei triggering heat instabilities and current jumps. A similar effect is known in vortex physics, where it was demonstrated that hotspots can cause vortex flow instabilities and corresponding voltage jumps.\cite{Xiao:1998} This indicates that the threshold electric properties of a two-phase medium can be dominated by statistically rare events of forming hotspots due to inherent inhomogeneity rather than average behavior and calls for a thorough study of this phenomenon. It has already been known that at the onset of MIT, the currents are concentrated along quasi-one-dimensional percolation-type optimal paths.\cite{Shashkin:1994,Okunev:1996} This stimulated extensive studies of statistical properties of optimal paths and the corresponding distribution of large currents.\cite{Chan:1989,Kahng:1987,Li:1987,Machta:1987,Shi:2013} However, properties and distribution of hot spots remain almost unexplored.

\begin{figure}[b]
	\begin{center}
	\includegraphics[width=8.5cm]{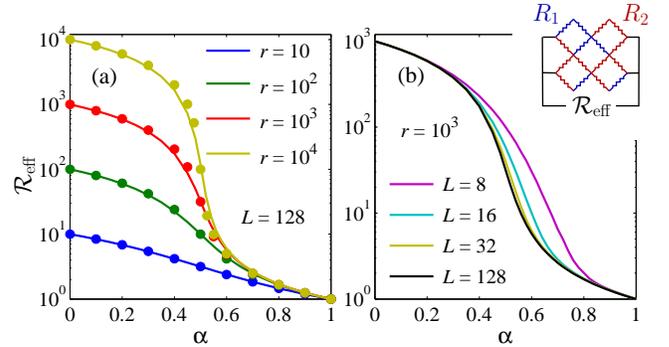}
	\vspace{-2mm}
	\caption{
		(a)~Effective resistance of a $L\times L$ diamond grid with two types 
		of resistors $R_1$ and $R_2$ as function of concentration $\alpha$ 
		of the first medium (the probability to find resistor $R_1$ in the grid), where 
		$L = 128$, $R_1 = 1$, and $R_2 = r R_1=10$, $10^2$, $10^3$, and $10^4$
		For comparison the effective resistance of a two-dimensional medium 
		with circular inclusions is shown by circles.
		(b)~Effective resistance of the grid as function of concentration 
		with $r = 10^3$ and different sizes, $L=8$, $16$, $32$, and $128$.
		Inset: $4\times 4$ diamond grid with two types of resistors, $R_1$ and $R_2$.
	}
	\label{fig:gridResistance}
	\end{center}
\end{figure}

\begin{figure*}[!t]
	\begin{center}
	\includegraphics[width=15.4cm]{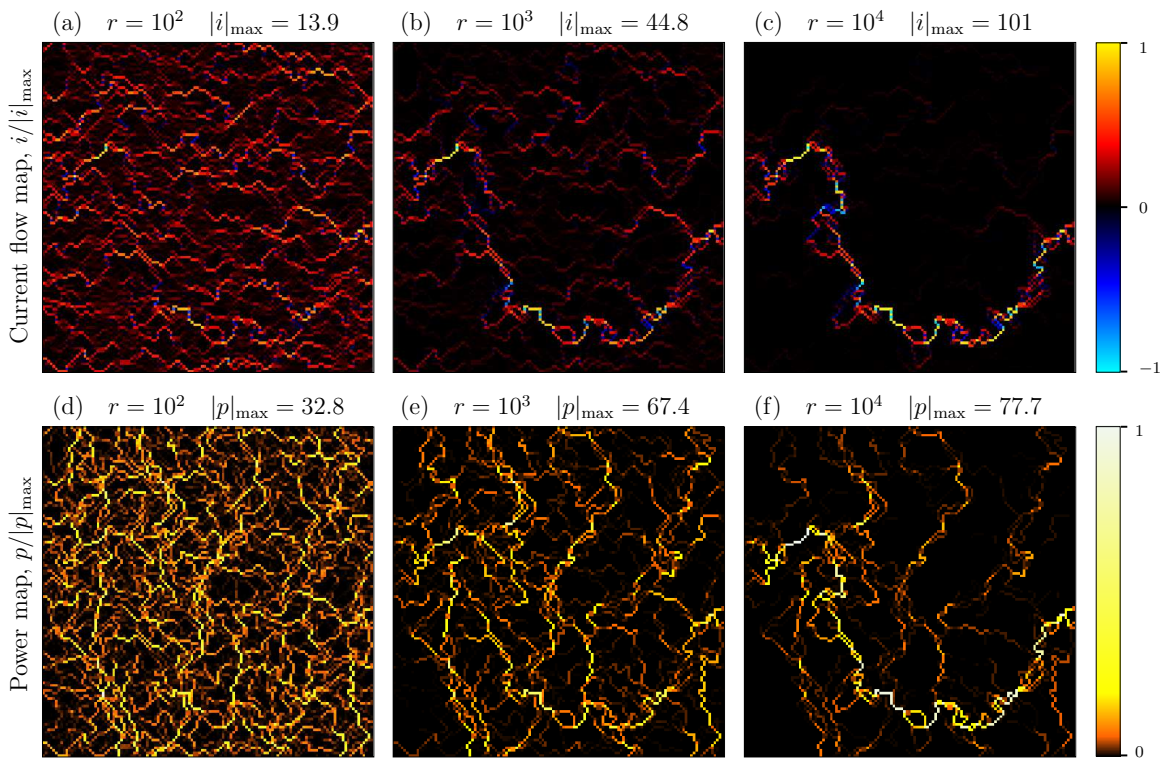}
	\vspace{-1mm}
	\caption{
		(a)--(c)~Current maps for a $L\times L$ resistor network 
		with $L=128$ and fixed distribution of the resistors $R_1$ 
		and $R_2$ at different ratios $r = 10^2$, $10^3$, and $10^4$.
		At volume fraction $\alpha = 0.5$ the current flows along 
		the optimal paths similar to those arising in the bond 
		percolation problem near the threshold (in a percolation 
		problem: $r = \infty$ and $L = \infty$). 
		The average size of current clusters oriented along the 
		applied current axis grows with increasing $r$.
		(d)--(f)~Power dissipation maps displaying the sites with 
		enhanced heat emission. One can see the correspondence 
		between the optimal paths map and the distribution of hotspots, 
		which mostly follow the optimal paths.
	}
	\label{fig:maps} 
	\end{center}
\end{figure*}

Here we address the question of the occurrence and spatial distribution of hot spots in a random two-dimensional two-phase medium. In Sec.~\ref{sec:model} we describe our model and approach, Sec.~\ref{sec:current} is devoted to the analysis of current and voltage to show the validity of our approach, in Sec.~\ref{sec:power} we give the statistical description of hot spots, and we finally summarize our results in Sec.~\ref{sec:conclusion}.

\section{Model
\label{sec:model}}

There have been extensive studies of two-phase media (TPM), which offer good models of structurally disordered systems, resulting in the fundamental and exact expression by Dykhne for the effective resistance of a TPM with equal concentrations $\alpha = 1/2$ of randomly distributed phases with resistivities $R_1$ and $R_2$ respectively:\cite{Dykhne:1970} $\mathcal{R}_{\rm eff} = \sqrt{R_1 R_2}$. This expression follows from the existence of a fixed point of the duality transformation interchanging phases 1 and 2,
$
	\mathcal{R}_{\rm eff}(\alpha, R_1, R_2)
	\mathcal{R}_{\rm eff}(1-\alpha, R_1, R_2) =
	R_1 R_2.
$
For $\alpha \neq 1/2$ the effective resistivity of the medium depends on the realization of disorder and is not of general nature as in the case of $\alpha = 1/2$. Considerable efforts have been made to extend Dykhne's result onto more general situations, in particular for the case of arbitrary concentrations\cite{Bulgadaev:2003a,Bulgadaev:2003b,Bulgadaev:2003c} and multi-component media. An alternative approach is based on constructing an effective medium approximation.\cite{Kirkpatrick:1973} Useful results were derived for a model system comprising spherical conducting inclusions embedded into the matrix with the different conductivity: the resulting effective resistance $\mathcal{R}_{\rm eff}$ of $D$-dimensional $N$-phase media is\cite{Landauer:1978}
$
	\sum_{k=1}^N \alpha_k
	[R_{k}^{-1} - \mathcal{R}_{\rm eff}^{-1}] /
	[R_{k}^{-1}+(D-1)\mathcal{R}_{\rm eff}^{-1}] = 0,
$ 
where $\alpha_{k}$ and $R_{k}$ are concentration and resistivity of the $k^{\rm th}$ component, $\sum_{k=1}^N \alpha_k = 1$, and $k = 1 \ldots N$. At $N = 2$ and $\alpha_1 = \alpha_2=1/2$ this expression reduces to Dykhne's result. For small deviations of randomly distributed resistivities~$R$ from their average $\langle R\rangle$, $|\langle R^{2}\rangle-\langle R\rangle^{2}| \ll \langle R\rangle^{2}$, the effective resistance becomes\cite{Herring:1960}
$ 
	\mathcal{R}_{\rm eff} =
	\langle R\rangle
	[1 - 
		2(\langle R^2\rangle - \langle R\rangle^2) /
		3\langle R\rangle^2 
	],
$
where $\langle \cdot \rangle$ denotes the spatial average. We will be using the asymptotic expressions for $\mathcal{R}_{\rm eff}$ for verification of our numerical approach, see, e.g., Fig.~\ref{fig:gridResistance}.

We will study the general case of two-phase media at arbitrary values of resistance ratio and their mutual concentration. To model a statistically isotropic TPM we employ a random rhombic/diamond two-component resistor network (TCRN) shown in the inset of Fig.~\ref{fig:gridResistance}(b). On each bond either a resistance $R_1$ or $R_2$ are placed with concentrations~$\alpha$ or $1-\alpha$, respectively. The network is confined between two electrodes, the source and the drain. The controlling parameters of the TCRN are the partial fraction $\alpha$, $0 \leqslant \alpha \leqslant 1$, of the resistance $R_1$ and the ratio of the respective resistances $r=R_1/R_2$. We assume $R_1 \geqslant R_2$, such that $r \geqslant 1$. Values of $r \gg 1$ and $\alpha$ near $1/2$ mimic to the situation near the MIT.\cite{Dubrov:1976} Thus, the overall system is characterized by a $L\times L$ resistor matrix $R_{nm}$ (elements $R_1$ or $R_2$), where~$L$ is a positive even integer proportional to the linear system size of the modeled TPM and $n, m = 1, \ldots, L$. A finite voltage~$\mathcal{V}$ is applied across the sample. The resulting currents in the system, $I_{nm}$, and the total resistance $\mathcal{R}_{\rm eff}$ of the system is found by using Kirchhoff's circuit laws. Numerically, we calculate the inverse of a $L^2 \times L^2$ sparse matrix with the $(2L-1)^2$ nonzero elements.

The dependence of the total effective resistance $\mathcal{R}_{\rm eff}$ on the concentration $\alpha$ is shown in Fig.~\ref{fig:gridResistance}(a) for $L = 128$ and different resistance ratios $r = R_2/R_1 = 10$, $10^2$, $10^3$, and $10^4$. The comparison of the different grid sizes $L=8$, $16$, $32$, and $128$ at $r = 10^3$ is given in Fig.~\ref{fig:gridResistance}(b). Our numerical results for $\alpha = 1/2$ reproduce Dykhne's result. For~$r$ deviating not too much from unity, the result of our calculations coincides with the effective resistance obtained using an analytical derivation for $|R_1 - R_2| \ll \langle R \rangle$ (not shown). The effective resistance for a continuous two-dimensional system with circular inclusions is depicted by circles in Fig.~\ref{fig:gridResistance}(a). For $r \lesssim 10^2$ it is in good agreement with the effective resistance of the TCRN; only for $r \gtrsim 10^3$ the computed effective resistance starts to differ noticeably from the circular inclusions case. 

\begin{figure}[!b]
	\begin{center}
	\includegraphics[width=8.5cm]{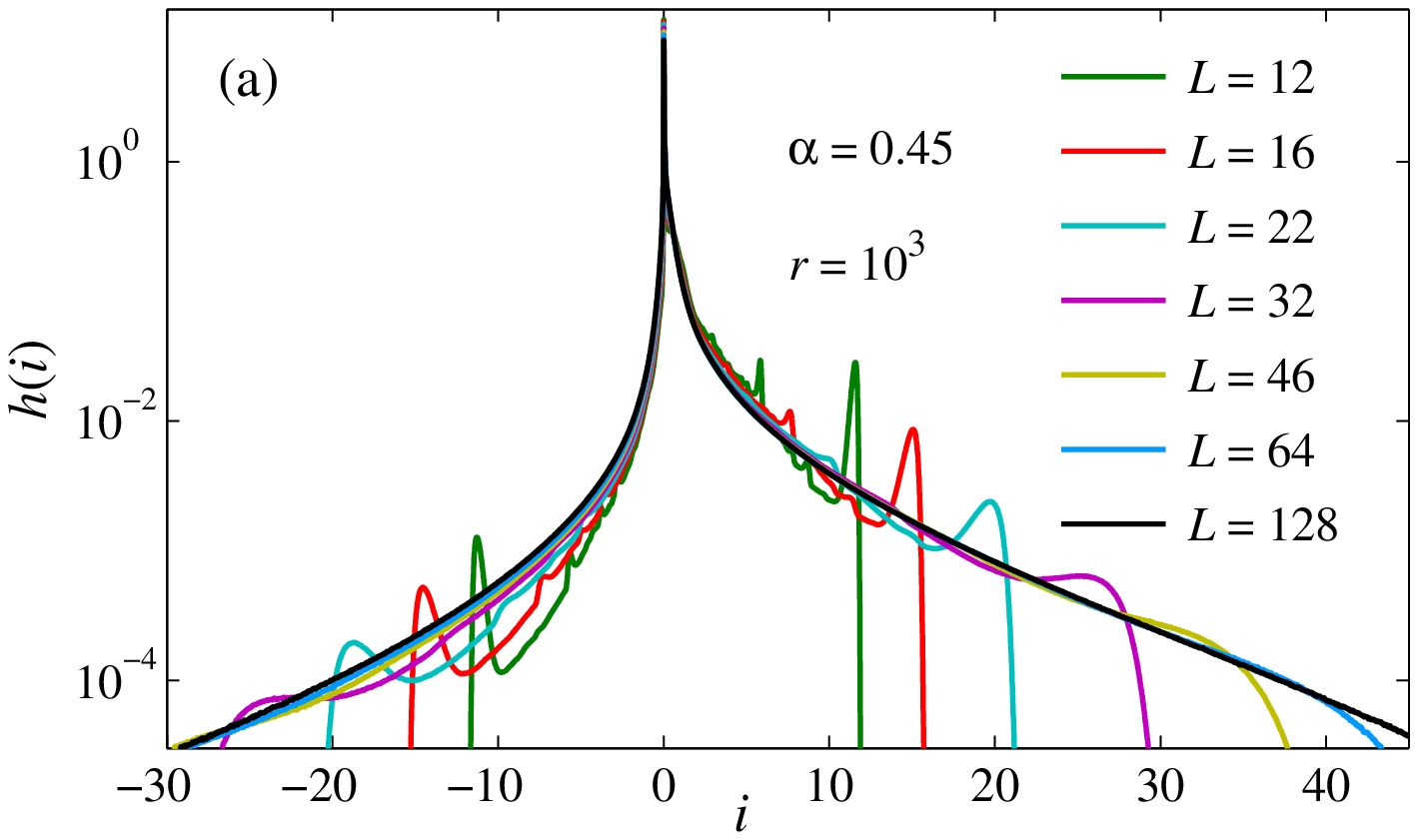} \\ \vspace{1.5mm}
	\includegraphics[width=4.25cm]{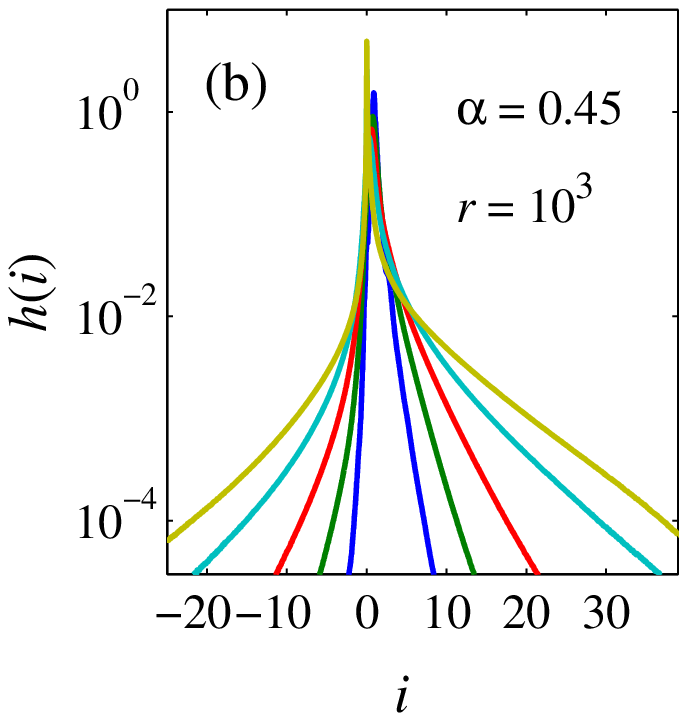}
	\includegraphics[width=4.25cm]{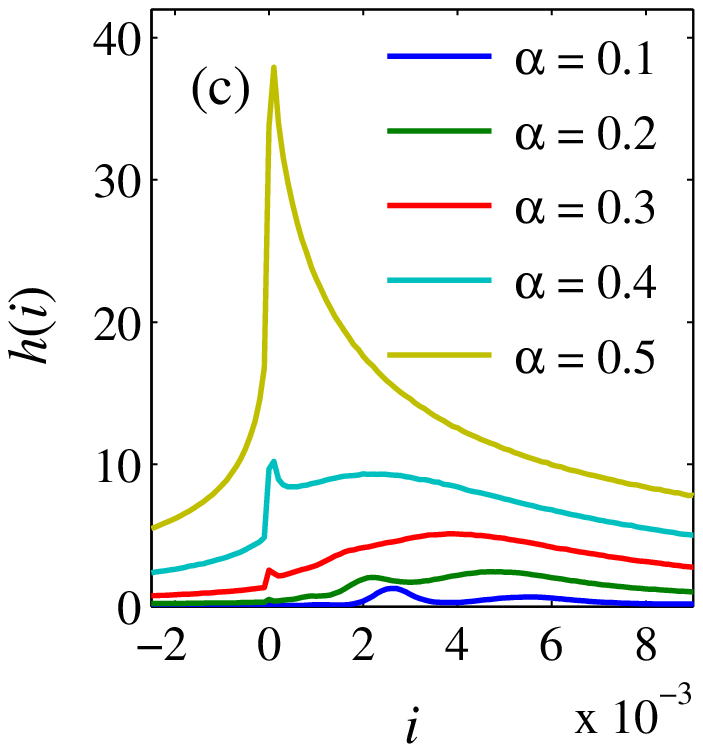}
	\\ \vspace{-2mm}
	\caption{
		(a)~PDF of the currents flowing through individual resistors, $h_{r,\alpha}(i)$,
		for volume ratio $\alpha = 0.45$, resistance ratio $r=10^3$, 
		and different system sizes~$L$.
		(b)~Current PDF for different volume ratios $\alpha$, $r=10^3$, 
		and $L = 128$ in semilogarithmic scale.
		(c)~Closeup of (b) in linear scale.
	}
	\label{fig:currentHist} 
	\end{center}
\end{figure}

\begin{figure}[!b]
	\begin{center}
	\includegraphics[width=8.5cm]{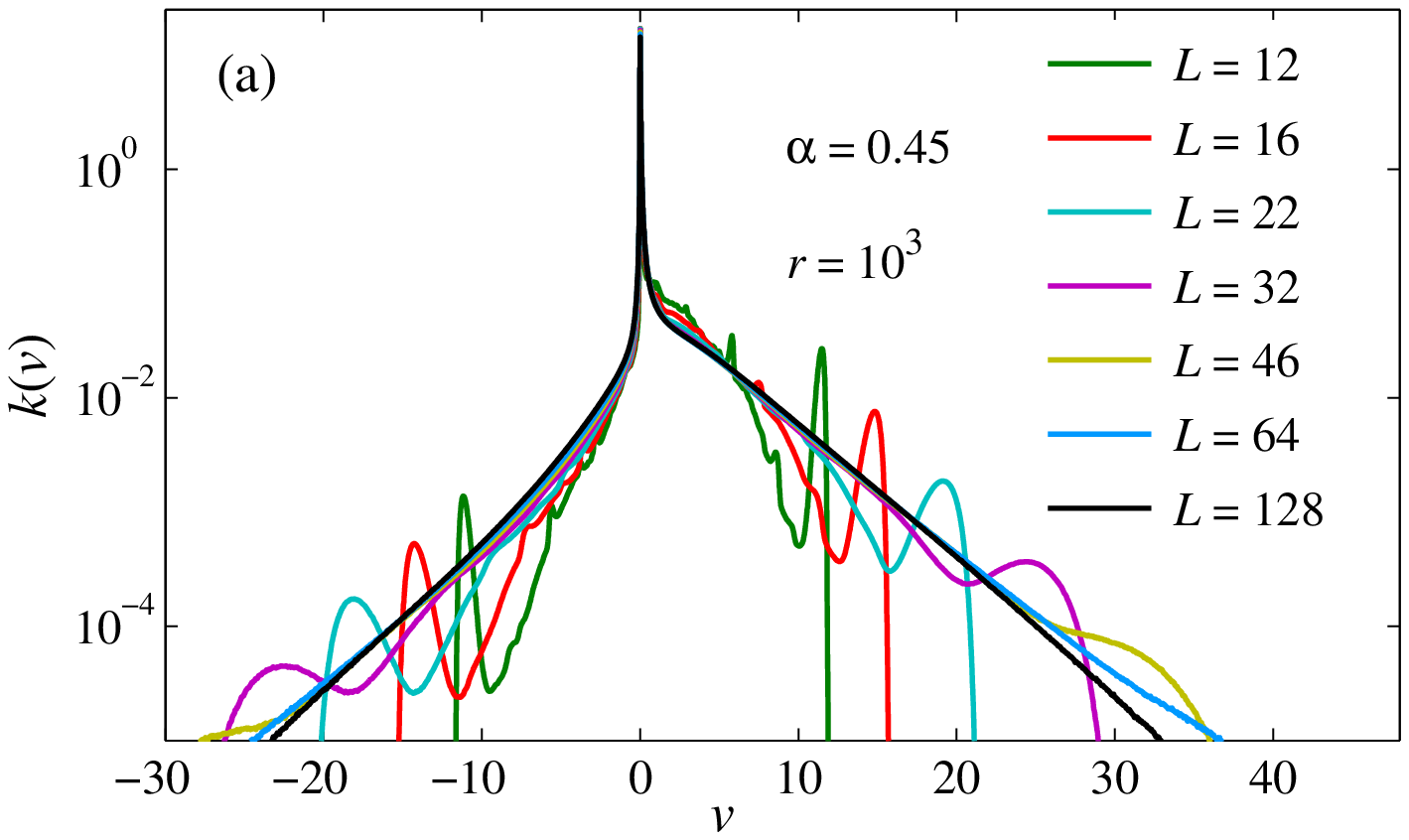} \\ \vspace{1.5mm}
	\includegraphics[width=4.25cm]{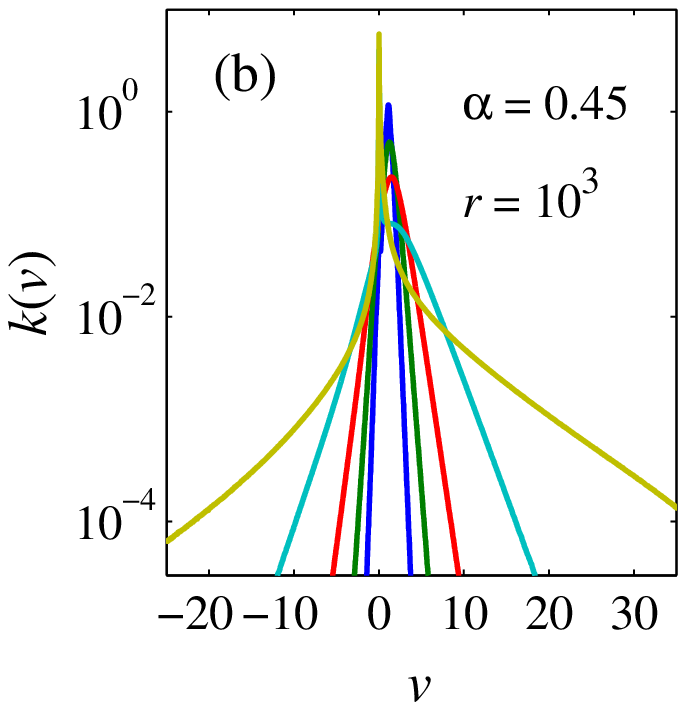}
	\includegraphics[width=4.25cm]{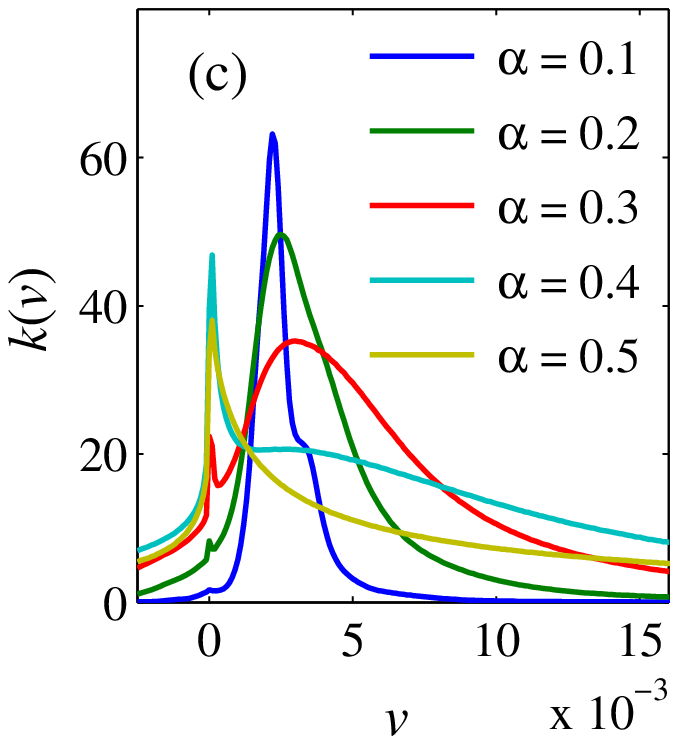} \\ \vspace{-2mm}
	\caption{
		(a)~Voltage PDF $h_{r,\alpha}(i)$ for volume ratio $\alpha = 0.45$, 
		resistance ratio $r=10^3$, and different system sizes~$L$.
		(b)~The same for different volume ratios $\alpha$, $r=10^3$, 
		and $L = 128$ in semilogarithmic scale.
		(c)~Closeup of (b) in linear scale.
	}
	\label{fig:voltageHist}
	\end{center}
\end{figure}

Figure~\ref{fig:maps} shows spatial distribution of current paths [Figs.~\ref{fig:maps}(a)--\ref{fig:maps}(c)] and power dissipation across particular bonds [Figs.~\ref{fig:maps}(d)--\ref{fig:maps}(f)]. Colors indicate normalized values of the current $i_{nm} = I_{nm} L \mathcal{R}_{\rm eff} / \mathcal{V}$ and power $p_{nm} = P_{nm} L^2 \mathcal{R}_{\rm eff} / \mathcal{V}^2$ (indices $n$ and $m$ will be omitted in some instances in the text below) for the same realization of the distribution of the resistances $R_1$ and $R_2$, but having different resistance ratios $r$. The system with equal volume fractions, $\alpha = 1/2$, corresponds to the percolation threshold $\alpha_{\rm c} = 1/2$ at $r = \infty$. For $r = 1$ (not shown) the spatial distribution of the current is uniform, as it should. With increasing $r$ the current flow starts to concentrate along favorable pathways; for $r=10^2$ the optimal paths are clearly seen, Fig.~\ref{fig:maps}(a). Upon further increase of $r$, only a few preferable paths survive [Fig.~\ref{fig:maps}(b)], until only a single critical backbone cluster at $r = 10^4$ dominates the current transfer [Fig.~\ref{fig:maps}(c)]. For large~$r$ this backbone cluster clearly coincides with that of the percolation problem near the percolation threshold. Remarkably, the spots of the enhanced power dissipation form cluster configurations nearly coinciding with current optimal paths. 

\begin{figure}[b]
	\begin{center}
	\includegraphics[width=8.5cm]{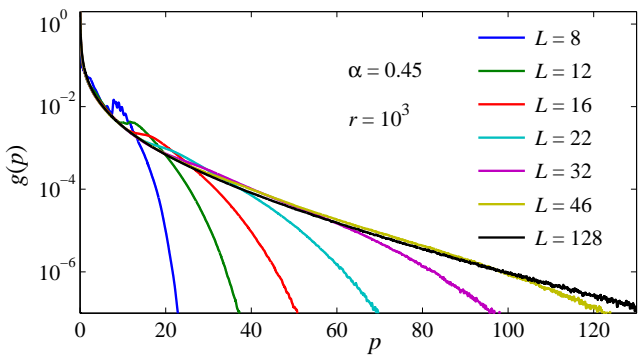}
	\vspace{-2mm}
	\caption{
		Probability density function (PDF) $g_{r,\alpha}(p)$ of the normalized 
		power emitted by individual resistors (`power distribution') for different 
		system sizes $L$ on semilogarithmic scale for $\alpha = 0.45$ and $r = 10^3$.
		The distribution $g_{r,\alpha}(p)$ demonstrates exponential behavior, 
		see Eq.~(\ref{eq:powerPDF}).
	}
	\label{fig:powerHistSize}
	\end{center}
\end{figure}

\begin{figure}[b]
	\begin{center}
	\includegraphics[width=8.5cm]{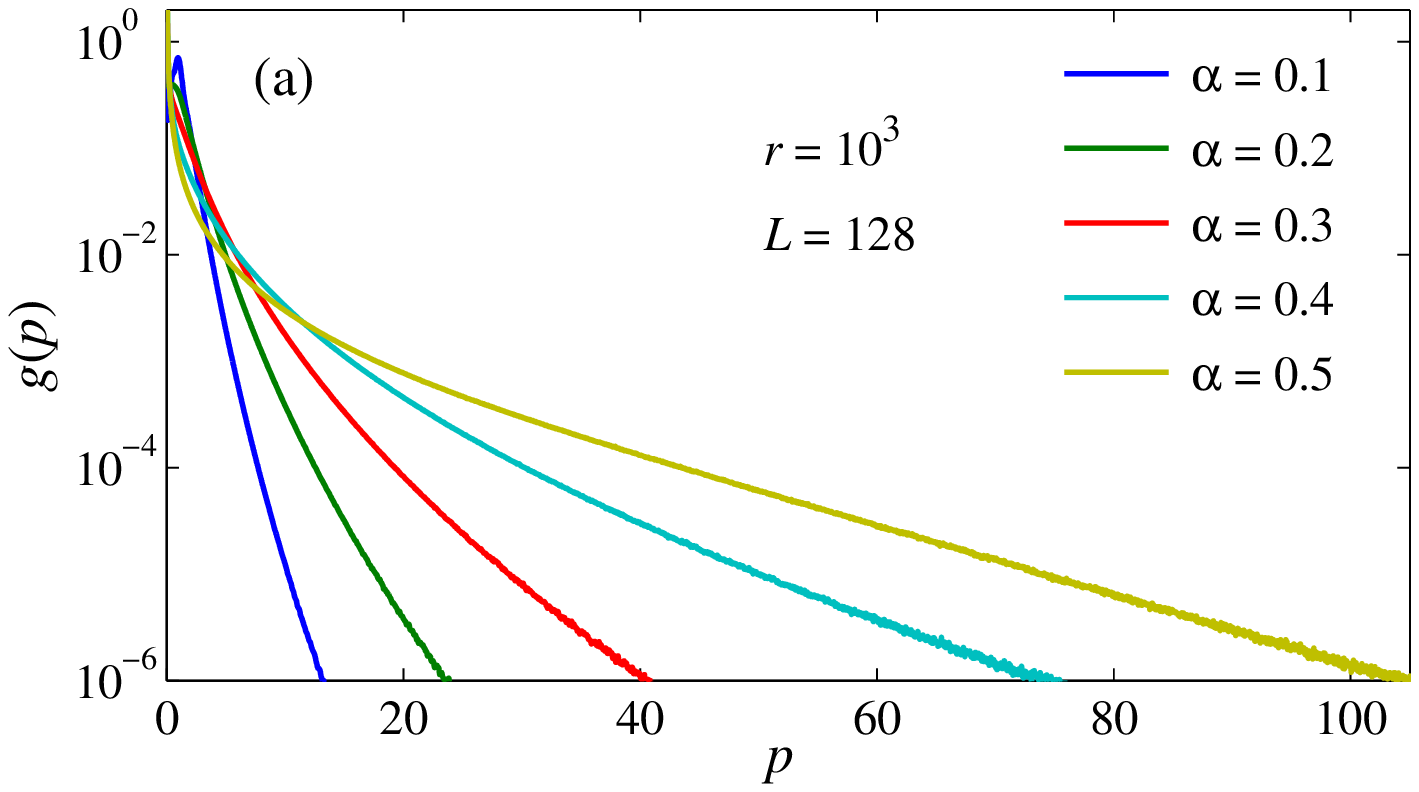} \\ \vspace{1mm}
	\includegraphics[width=4.25cm]{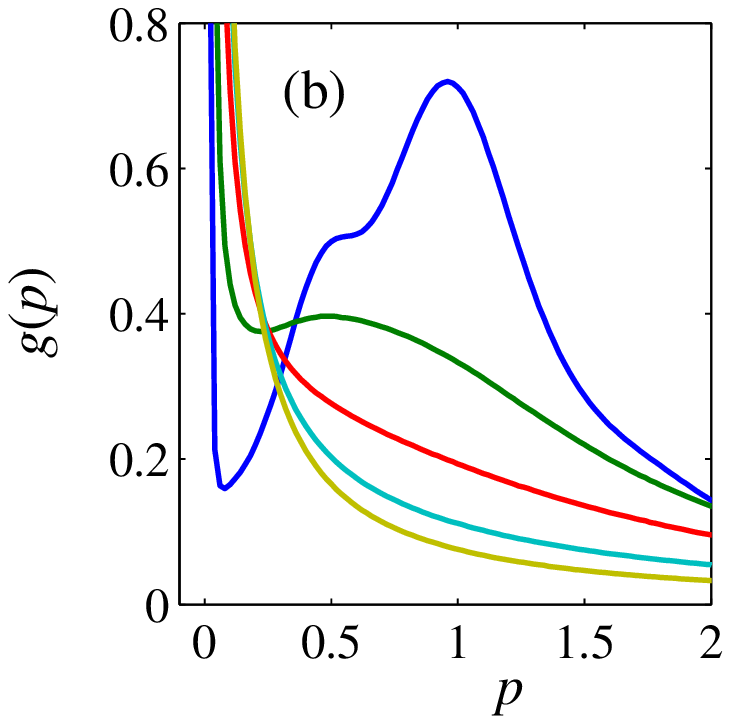}
	\includegraphics[width=4.25cm]{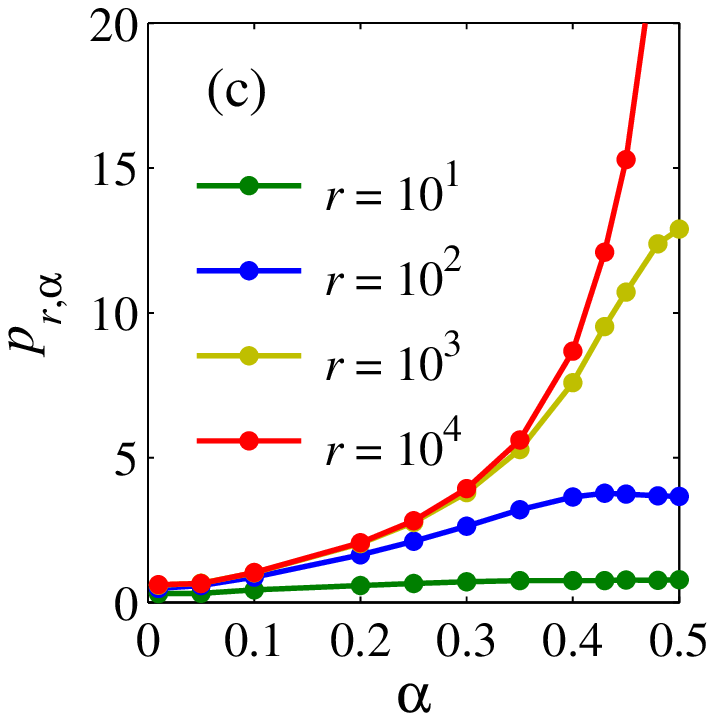} \\ \vspace{-2mm}
	\caption{
		(a)~Power PDF $g_{r,\alpha}(p)$ for different concentrations 
		$\alpha$, system size $L = 128$, and resistance ratio~$r = 10^3$
		in semilogarithmic scale.
		(b)~Closeup in linear scale, only the top parts of the curves are 
		shown, where some of them develop nonmonotonic behavior.
		(c)~The slope $p_{r,\alpha}$ of the exponential tail of the PDF 
		shown in panel~(a) as a function of concentration $\alpha$ 
		for different resistance ratios~$r$.
	}
	\label{fig:powerHistConcentration}
	\end{center}
\end{figure}

\section{Current and voltage distribution
\label{sec:current}}

The distribution of the currents flowing through individual resistors and the corresponding voltage drops has been studied in great detail.\cite{Arcangelis:1986,Kolek:1996,Shi:2013} We use these known results to validate our model and therefore represents the first step for the investigation of the dissipated power distribution. To this end, we introduce dimensionless quantities $i_{nm} = I_{nm} L \mathcal{R}_{\rm eff} / \mathcal{V}$ and $v_{nm} = V_{nm} L / \mathcal{V}$, where $V_{nm} = I_{nm} R_{nm}$. The distributions of current and voltage follow, as expected, exponential behavior.\cite{Li:1987,Shi:2013} 
Current and voltage PDFs are shown in Figs.~\ref{fig:currentHist} and \ref{fig:voltageHist}, respectively.

In Fig.~\ref{fig:currentHist}(a) the current PDF $h_{r,\alpha}(i)$ for resistance ratio $r = 10^3$, concentration $\alpha = 0.45$, and different system sizes $L$ is plotted in semilogarithmic scale. One can see a large peak at $i = 0$, which corresponds to inactive zones with almost zero currents. In small systems one observes prominent peaks at $|i| \sim 10-20$, which usually corresponds to the current flowing through optimal paths as shown in Fig.~\ref{fig:maps}(b). For the absolute values of the current larger then its peak value, the PDF rapidly drops, i.e., one cannot find currents considerably exceeding the average one in small systems. With increasing system size $L$, the peak in the current PDF shifts towards larger absolute values of the current and size effects vanish. The envelope curve for $L = 128$ demonstrates size-independent exponential behavior for currents shown in the plot. The dependence on the concentration $\alpha$ is shown in Figs.~\ref{fig:currentHist}(b) and \ref{fig:currentHist}(c) for $r = 10^3$ and $L = 128$ in semilogarithmic and linear scales correspondingly. Inspecting the current distribution in a large system with resistivity ratio $r=10^3$, one observes an interesting feature: an appreciable fraction of resistors (bonds) with a significant number of resistors, having current flowing in the direction opposite to globally applied voltage. This is reflected in the highly meandering character of the optimal current paths and may have far reaching consequences for {\it superconducting} random networks in the fluctuation regime. Namely, one can expect that under the conditions, where non-dissipative currents are possible, close supercurrent loops will be appearing. These supercurrent loops mark positions of spontaneously generated vortices due to disorder.

The probability density function for voltage drops $k_{r,\alpha}(v)$ for different system sizes is plotted in Fig.~\ref{fig:voltageHist}(a). The voltage PDF has exponential tails as well, as shown by the envelope curve for $L = 128$. The dependence of the voltage PDF $k_{r,\alpha}(v)$ on the concentration $\alpha$ is shown in Fig.~\ref{fig:voltageHist}(c) and \ref{fig:voltageHist}(d).

\section{Power loss distribution and hot spots
\label{sec:power}}

Now, we turn to the main topic of this paper, the investigation of the distribution of the locally dissipated power $P_{nm} = I_{nm}^2 R_{nm}$. In order to determine the probability density function (PDF) of the power dissipated at individual resistors, we first construct a map of the dissipated power analogous to the map of the current distribution. Figures~\ref{fig:maps}(d)--\ref{fig:maps}(f) show maps of the dissipated power in the same system and with same ratios~$r$ as in Figs.~\ref{fig:maps}(a)--\ref{fig:maps}(c). We use the natural normalization $p_{nm} = P_{nm} / \langle P\rangle$, where $\langle P\rangle = \mathcal{V}^2 / \mathcal{R}_{\rm eff} L^2$ is the average power from individual resistors (or the power emitted from a single resistor in the homogeneous case). At a smaller ratio $r$, the dissipated power is distributed nearly uniformly and is only slightly modulated by small amplitude fluctuations, see Fig.~\ref{fig:maps}(d). Clusters consisting of the `hottest spots paths' become more pronounced upon increasing $r$, similar to clusters in percolation networks. When the correlation length of this cluster, the linear size of a cell of connected hotspots, becomes comparable to the system size $L$, only the one last critical `hot path' that traverses the system remains. Note, the configuration of the hotspot clusters follows those of the current-carrying clusters, which carry the most of the current.

To carry out the quantitative analysis we consider the dimensionless power $p$ emitted by individual resistors in the network, specifically its PDF $g_{r,\alpha}(p)$. It has the following symmetries 
\begin{subequations}
\begin{eqnarray}
	& g_{r,\alpha}(p) = g_{r,1-\alpha}(p),
	\label{eq:powerSymm1} \\
	& g_{r,\alpha}(p) = g_{1/r,\alpha}(p) 
	\label{eq:powerSymm2}.
\end{eqnarray}
\end{subequations}
The size dependence of the PDF $g_{r,\alpha}(p)$ is presented in Fig.~\ref{fig:powerHistSize} in semi-logarithmic scale. Those results are averaged over $3\times 10^5$ realizations of resistor distributions with $L = 128$, which correspond to $5\times 10^9$ individual resistors. One sees that at small system sizes the probability distribution density for finding an element emitting power~$p$, is non-universal and size-sensitive. Upon increasing the size~$L$, an universal PDF develops as an envelop curve converging, as $L\to\infty$, to the generic size-independent PDF for the TPM for given~$\alpha$ and $r$, having exponential tails:
\begin{equation}
	g_{r,\alpha}(p) \propto \exp(-p/p_{r,\alpha}).
	\label{eq:powerPDF}
\end{equation}
The power decays over the characteristic scale $p_{r,\alpha}$, which defines the slope of the PDF tail and depends on~$\alpha$ and~$r$.

\begin{figure}
	\begin{center}
	\includegraphics[width=8.5cm]{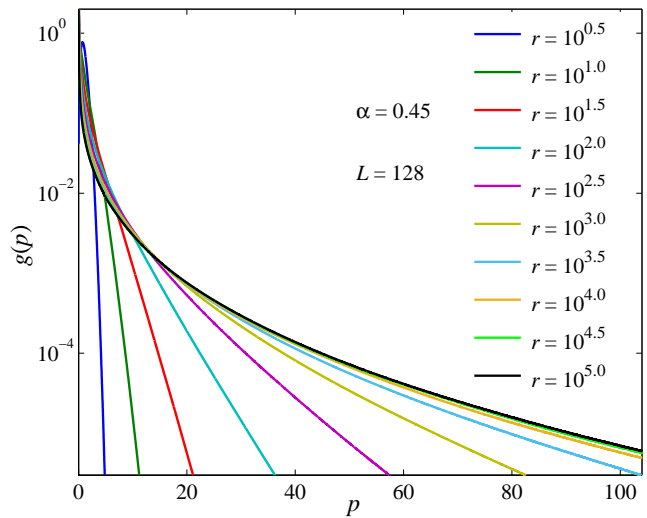}
	\vspace{-2mm}
	\caption{
		Power PDF $g_{r,\alpha}(p)$ for different resistance ratios~$r$, 
		concentration $\alpha = 0.45$, system size $L = 128$. 
		Longer tail corresponds to larger~$r$.
	}
	\label{fig:powerHistRatio} 
	\end{center}
\end{figure}

Next, we study the dependence of the PDF $g_{r,\alpha}(p)$ on the concentration~$\alpha$. The results are shown in Fig.~\ref{fig:powerHistConcentration}(a) as PDFs for different $\alpha$ from $0.1$ to $0.5$ [the PDFs in the range from $0.5$ to $0.9$ are identical due to Eq.~(\ref{eq:powerSymm1})]. Again, we observe exponential behavior for all concentrations. The PDF tail becomes `longer' for larger $\alpha$ in the range $[0 \ldots 1/2]$ and the decay parameter $p_{r,\alpha}$ increases with increasing $\alpha$ and $r$. The corresponding dependence of $p_{r,\alpha}$ on~$\alpha$ is shown in Fig.~\ref{fig:powerHistConcentration}(c) for different ratios~$r$. A closeup of the PDF $g_{r,\alpha}(p)$ near small $p$ in linear scale is shown in Fig.~\ref{fig:powerHistConcentration}(b). One can see a peak at $p = 1$ for small disorder $\alpha = 0.1$, which remains from the homogeneous distribution ($\alpha = 0$), where the PDF is a single $\delta$-function, $g_{r,\alpha}(p)=\delta(1-p)$. With increasing disorder towards $\alpha = 1/2$ the number of conducting paths decreases and the peak at $p = 1$ disappears.

A set of PDFs $g_{r,\alpha}(p)$ for different ratios $r$ are shown in Fig.~\ref{fig:powerHistRatio}. 
Larger $r$ corresponds to longer tails, see Fig.~\ref{fig:powerHistConcentration}(c).

\section{Conclusion
\label{sec:conclusion}}

We have studied statistical properties of conducting paths in two-phase media, which are modeled as random resistor networks with two different resistance values. We found that in the limit of a high ratio of the resistances, which corresponds to strongly disordered two-dimensional media, the transport across the random resistor network is dominated by percolation-like optimal paths. We confirmed that the distribution of the currents carried through these paths follows exponential statistics. We investigated statistical properties of the hotspot distribution governing the dissipation of the network and found that the distribution function of the locally dissipated power is exponential as well. An important implication of the obtained results is that heating instabilities occur exponentially more often in strongly disordered films than what could have been expected from Gaussian behavior.

A deep reason for the origin of the exponential distribution is that the optimal current paths are formed in such a way that Joule losses were minimal. This selection by optimum is realized on every spatial scale and results in extreme value statistics for distribution of relevant parameters governing dissipation. In conjunction with the percolation-like character of the optimal paths this should give rise to
exponential distribution of the resulting Joule losses.

Our results obtained in the classical limit for resistive random conducting media can be used as a reference for future studies of random superconducting films exhibiting a disorder-driven superconductor-insulator transition. Indeed, in these films the superconducting critical temperature has noticeable spatial variations. Therefore, a two-phase description of these films endowed with disorder-induced random distribution of critical temperatures near the insulating/normal state, could be used within the framework of a Ginzburg-Landau description. One can expect the appearance of effective `vortex traps,' which can serve as nuclei for the phase separation near the superconducting transition, where superconducting paths form a meandering backbone cluster.

Furthermore, results from the two-phase media model can serve as input for a full quantum calculation of a Josephson junction array describing superconducting films. In practice one can restrict the calculation for a two-dimensional or three-dimensional system in the vicinity of the percolation threshold to an effectively one-dimensional system with the same two-phase distribution as in the backbone cluster.

\subsection*{Acknowledgments}

We are delighted to thank I. S. Aronson for useful discussions. The work was supported by the Scientific Discovery through Advanced Computing (SciDAC) program funded by U.S. Department of Energy, Office of Science, Advanced Scientific Computing Research and Basic Energy Sciences, and by the U.S. Department of Energy, Office of Science, Materials Sciences and Engineering Division. The work of T.I.B. was partly supported by the Program Quantum Mesoscopic and Disordered Systems of the Russian Academy of Sciences, the Russian Foundation for Basic Research (Grant No. 12-02-00152), and the Ministry of Education and Science of the Russian Federation.

\end{document}